\documentclass{article}
\usepackage{spconf,amsmath,graphicx,epsfig,subfigure,enumitem}

\title{BACKGROUND--TRACKING ACOUSTIC FEATURES FOR GENRE IDENTIFICATION OF BROADCAST SHOWS}
\name{Oscar Saz, Mortaza Doulaty, Thomas Hain\thanks{This work was supported by the EPSRC Programme Grant EP/I031022/1 (Natural Speech Technology).}}
\address{Speech and Hearing Group, Department of Computer Science, University of Sheffield, UK}
\begin{document}
%
\maketitle
\begin{abstract}
This paper presents a novel method for extracting acoustic features that characterise the background environment in audio recordings.
These features are based on the output of an alignment that fits multiple parallel background--based Constrained
Maximum Likelihood Linear Regression
transformations asynchronously to the input audio signal.
With this setup, the resulting features can track changes in the audio background like
appearance and disappearance of music, applause or laughter, independently of the speakers in the
foreground of the audio.
The ability to provide this type of acoustic description in audiovisual data has many potential
applications, including automatic classification of broadcast archives or improving automatic transcription and subtitling.
In this paper, the performance of these features in a genre identification task in a set of 332 BBC shows is explored.
The proposed background--tracking features outperform short--term Perceptual Linear Prediction features
in this task using Gaussian Mixture Model classifiers (62\% vs 72\% accuracy).
The use of more complex classifiers, Hidden Markov Models and Support Vector Machines,
increases the performance of the system with the novel background--tracking features to 79\% and 81\% in accuracy respectively.
\end{abstract}

\section{Introduction}

The media domain presents many opportunities for the application of speech technologies. With audiovisual data growing larger and larger every day
due to digital television, social media and on--line streaming there is a great need for performing automatic processing of this type of data.
Possible applications include automatic transcription and subtitling, classification of audiovisual archives and acoustic information
retrieval. Further research in this area is being also pushed by initiatives like the MediaEval Benchmarking for Multimedia Evaluation \cite{MediaEval},
which covers several of these tasks in the multimedia domain. The technologies required cover the whole range of speech
technologies: Automatic Speech Recognition (ASR); speaker identification; diarisation; identification of acoustic events; etc.

The ability of automatically detecting the genre of a broadcast show falls within the set of potential applications of speech technologies that
could become useful within the media domain. While genres are subjective divisions usually defined depending on the content of
the show, shows belonging to the same genre will share similar acoustic conditions that can be detected using automatic speech processing.
In this context, multimodal approaches, merging features from audio and video processing,
have been very commonly used \cite{Montagnuolo07,Montagnuolo09,Ekenel13,Mironica13} and have consistently provided results above 90-95\% accuracy.
Regarding the type of acoustic features used, from the early works the focus of research
has been on the use of short--term features \cite{Liu98}, including Mel--Frequency Cepstral Coefficient (MFCC) features \cite{Roach01}.
A full evaluation of the use of
MFCCs and Gaussian Mixture Model (GMM) classifiers
across 3 different test sets achieved 86\% in a RAI dataset, 78\%
in a Quaero dataset and 58\% in a YouTube dataset \cite{Ekenel13}. Using also MFCCs and GMMs, other authors achieved 94\% accuracy
in the RAI dataset when processing whole shows and 82\% on segments as short as 6 seconds \cite{Kim13}.

The different performances across sets indicate that short--term spectral features present
solid classification abilities, but are not robust in heterogeneous and complex datasets.
MFCCs, as well as Perceptual Linear Prediction (PLP) features \cite{Hermansky90}, represent the short--term characteristics
of speech, like the spectral properties of phonemes and speakers, but are not designed to characterise long--term
properties of audio. This could explain why, in homogeneous datasets,
where shows and speakers might often recur, like episodes from the same TV series or broadcast news programmes,
MFCCs performed outstandingly. A solution to this was proposed using Factor Analysis (FA) to extract
factors related to the genre, achieving 50\% improvement over the use of MFCC features
on Internet videos \cite{Rouvier09}.

Other approaches to this task \cite{Lee10,Castan13}
aim to identify specific audiovisual events that can be used as semantic blocks to understand the narrative of the
overall show or video. However this is a more complex task, due to the need to identify very subtle events, and its performance still
does not match the works previously mentioned in genre identification.

The work in this paper aims to provide a novel set of long--term background--tracking features that can perform a more natural description of
the type of acoustic background present, also tracking its temporal variations.
In order to have robust genre classification abilities, these features should be able to represent different background
conditions that can characterise shows, like studio recordings, outdoor noises, applause, laughter, different types of music, etc.
On the other hand, to ensure generalisation in the genre classification task, the features should factor out the influence of the speaker and the foreground.
The proposal explored in this work arises from the output of an asynchronous factorisation of background and speaker
with feature transformations, previously used in an ASR task \cite{Saz13}.


This paper is organised as follows: Section \ref{sec:finger} will present the audio processing system used to extract the background--tracking features
from audio files. Section \ref{sec:setup} will describe the experimental setup designed to perform
genre identification in a set
of broadcast shows from the BBC. Finally, Sections \ref{sec:results} and \ref{sec:conclusion} will present the results and
conclusions of this work.

\section{Background--Tracking Features}
\label{sec:finger}

In \cite{Saz13}, a novel method was presented to perform asynchronous factorisation of background and speaker in ASR tasks.
This method relied in using a set of Constrained Maximum Likelihood Linear Regression (CMLLR) transformations \cite{Gales96}
characterising different possible background conditions
that were switched asynchronously in the training and decoding process.
As a byproduct, applying this set of background transformations
asynchronously on a given audio segment will yield a sequence of states that will indicate
which CMLLR transform was applied in each frame and, hence, which corresponding background was considered to be more likely.

The first step in order to extract the proposed background--tracking features is to use a previously trained Hidden Markov Model (HMM) to align the input
audio data to its transcription, or to the output of a previous decoding if the transcription is not available, using a set
of asynchronous CMLLR transformations trained to represent different background conditions. 
The sequence of transformations applied in the best path from the alignment can be written into a vector $x=\{x(0),x(1),...,x(n),...,x(N-1)\}$, with $N$ being the length
of the input audio signal in frames and each value $x(n)$ given by the index assigned to each background CMLLR transformation
from a fixed set of values $\{0,1,...,t,...,T-1\}$, where $T$ is
the total number of background CMLLR transforms. Indicator functions $c_t(n)$, as defined in Equation \ref{eq:bin}, can be used to identify whether
the value of $x(n)$ is $t$ or not.

\begin{equation}
  \label{eq:bin}
  c_t(n)= \left\{ \begin{array}{rl}
 1 &\mbox{ if $x(n)=t$} \\
  0 &\mbox{ otherwise}
       \end{array} \right.
\end{equation}

The new feature vector proposed in this work is denoted as $v(m)$ and can be calculated as the moving average of the indicator functions $c_t(n)$ over a span
of $P$ of the original frames. This new vector has a length of $M=N/P$ and a dimension of $T$,
being formed by all the values $v_t(m)$, computed as in Equation \ref{eq:acc}, generating $v(m)=\{v_0(m),v_1(m),...,v_t(m),...,v_{T-1}(m)\}$.

\begin{equation}
  \label{eq:acc}
  v_t(m)=\frac{1}{P}\sum\limits_{p=0}^{P-1}c_t(m*P+p)
\end{equation}

\begin{figure}[t]
\centering
\epsfig{figure=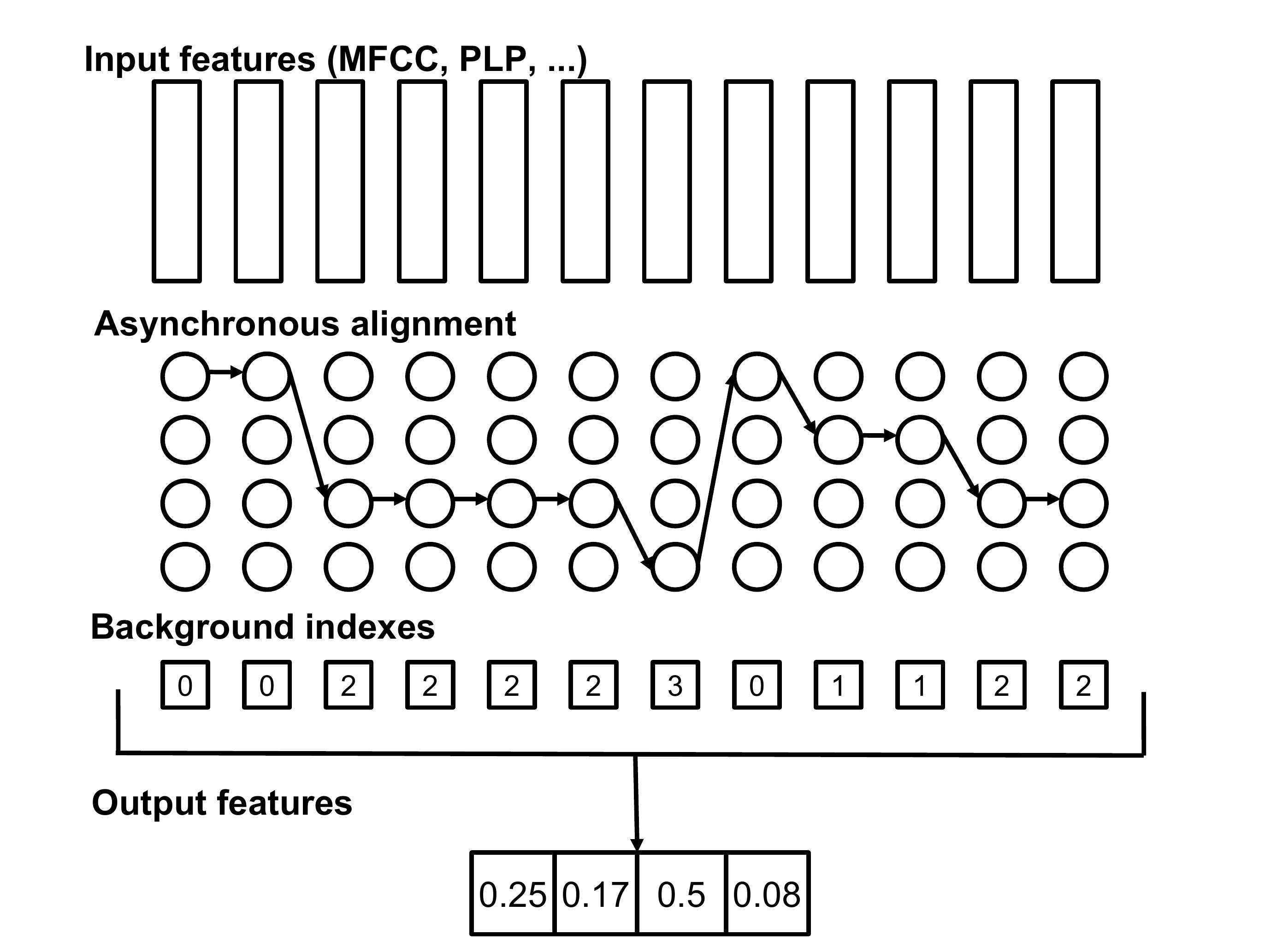,width=80mm}
\caption{{\it Background--tracking feature extraction.}}
\label{fig:finger}
\end{figure}

A graphical description of how this process is done can be seen in Figure \ref{fig:finger}. In this example, there are $T=4$ possible background transformations,
and values are aggregated every $P=12$ frames of the original input vector $x(n)$ generated as output of the asynchronous alignment.

\begin{figure*}[t]
\centering
\subfigure[News programme (Broadcast news)]{
    \epsfig{figure=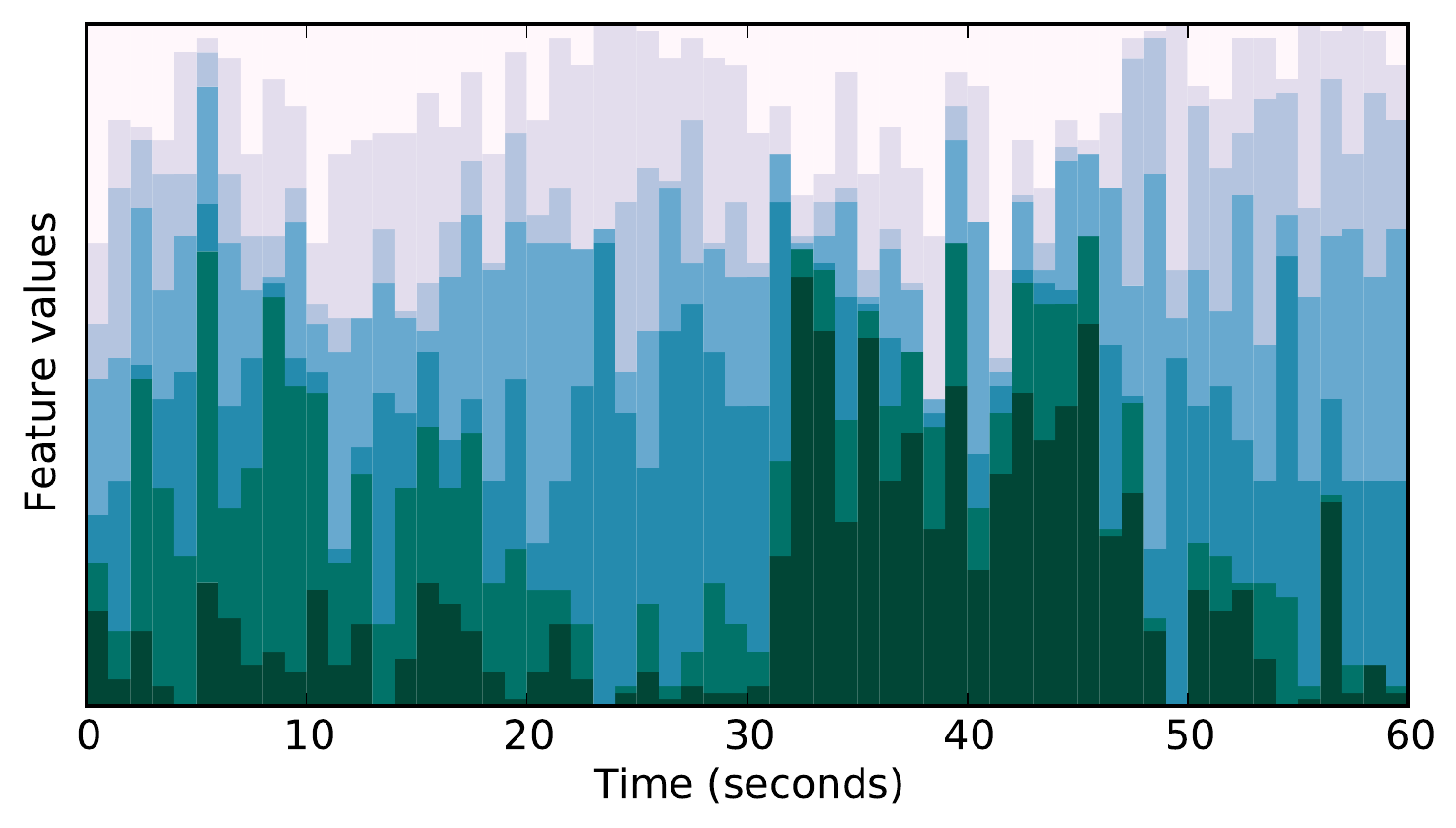,width=70mm}
    \label{fig:news}
}
\subfigure[Events programme (Live music show)]{
    \epsfig{figure=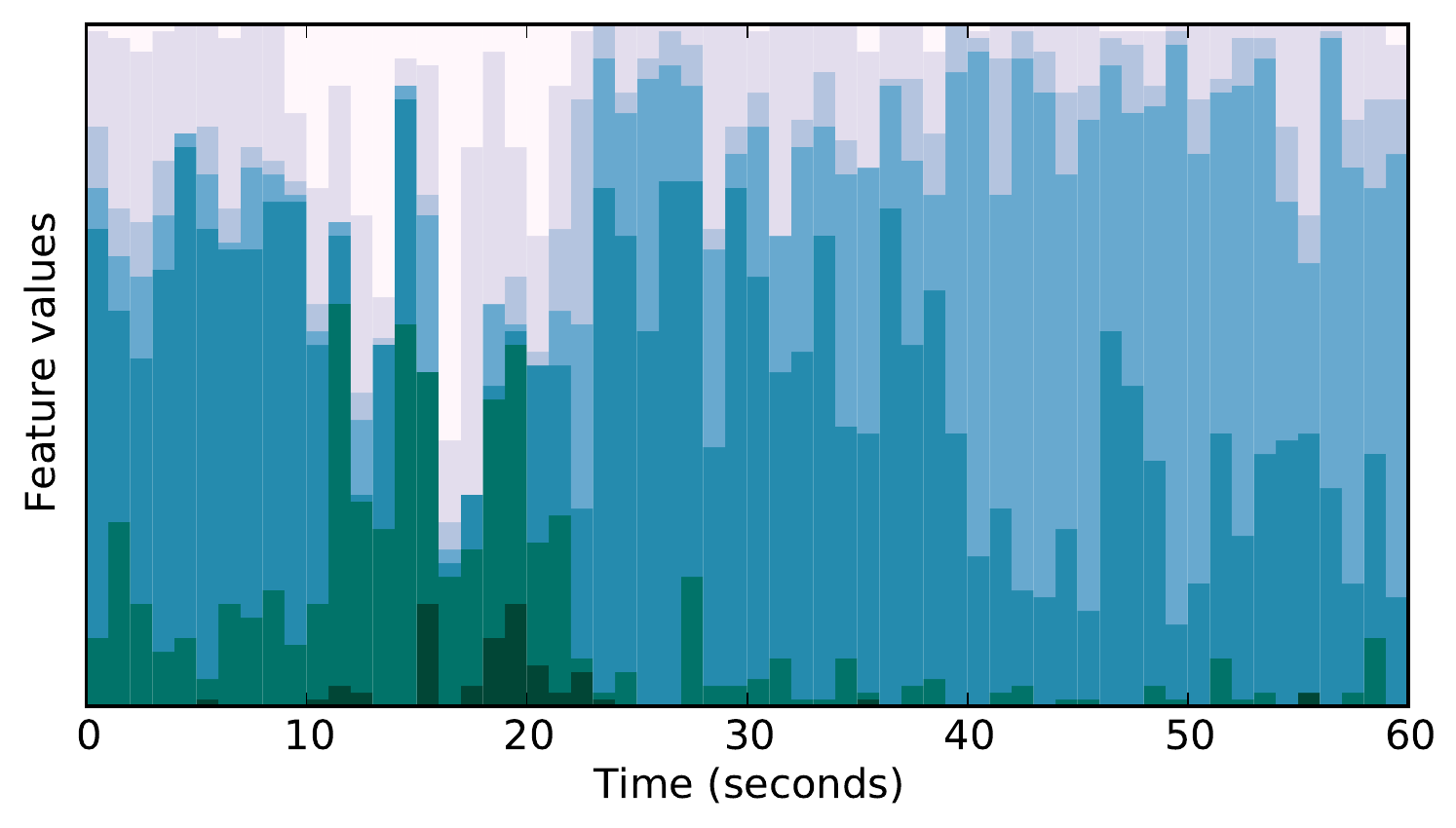,width=70mm}
    \label{fig:events}
}
\subfigure[Documentary programme (History show)]{
    \epsfig{figure=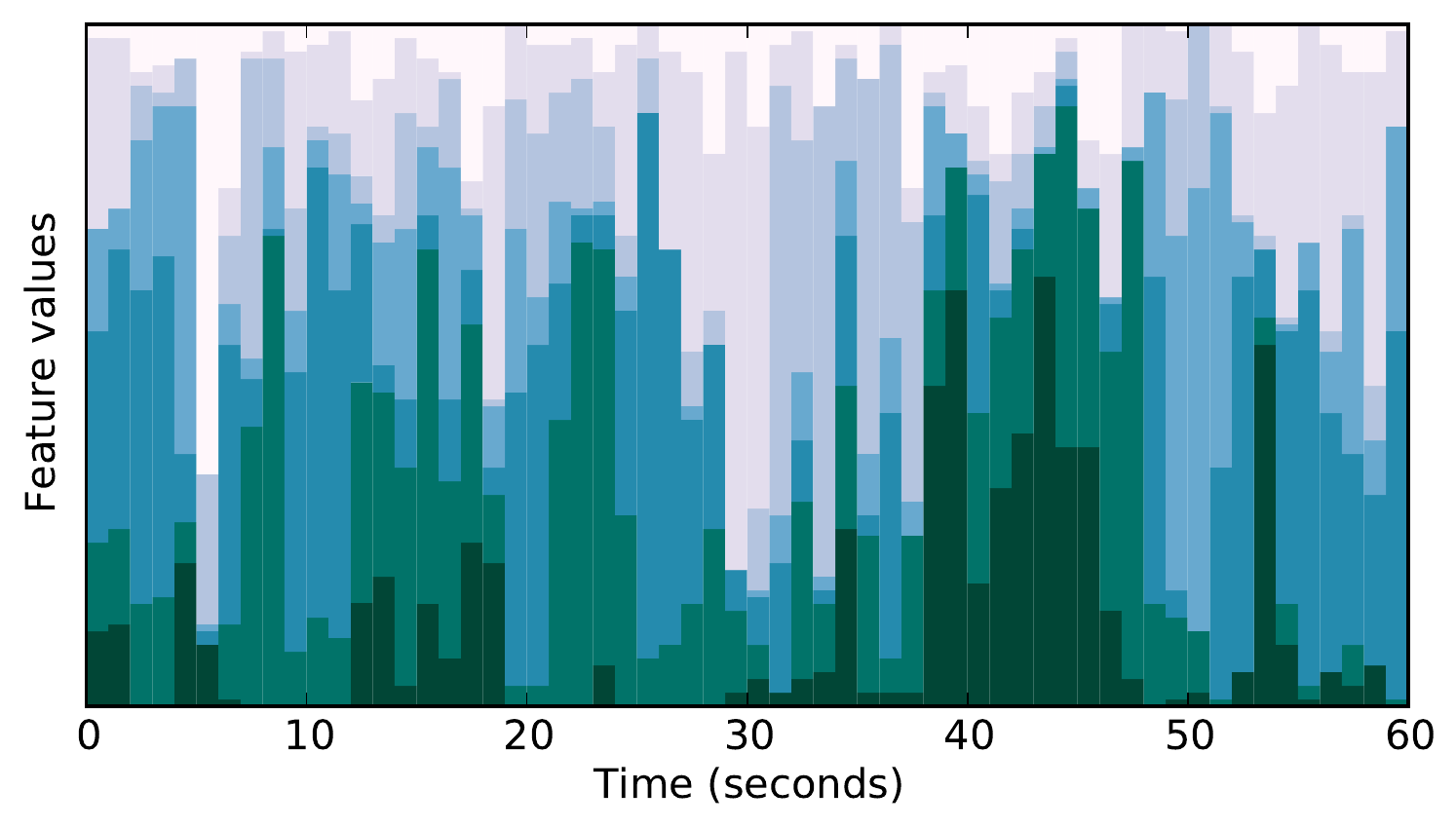,width=70mm}
    \label{fig:documentary}
}
\subfigure[Comedy programme (Light entertainment)]{
    \epsfig{figure=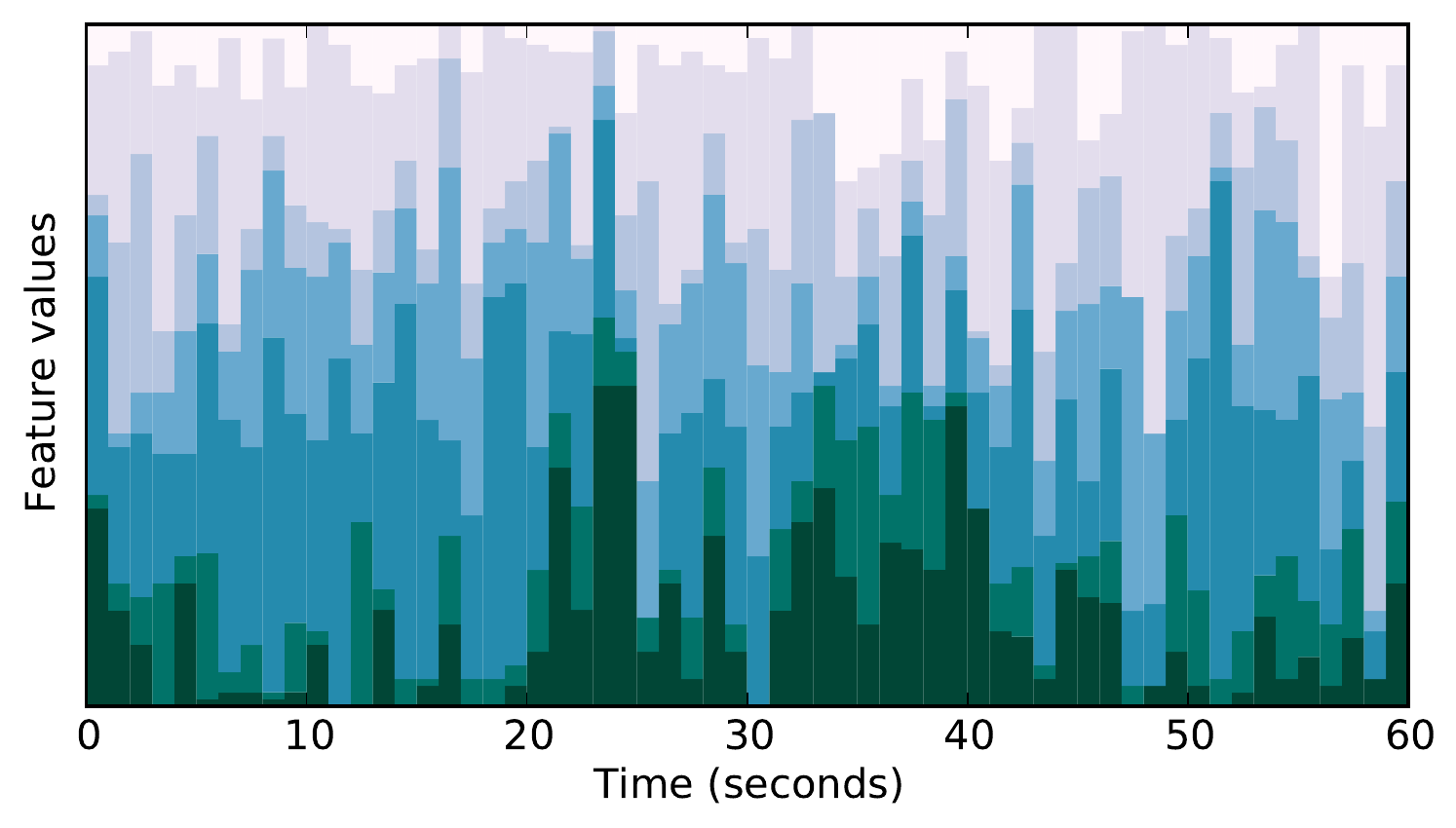,width=70mm}
    \label{fig:comedy}
}
\caption{{\it 1--minute samples of background--tracking features for different shows}}
\label{fig:samples}
\end{figure*}

\section{Experimental setup}
\label{sec:setup}

The experiments for the evaluation of the proposed background--tracking features were done in a set of 332 shows, totalling 231 hours,
broadcast by the BBC during the first
week of May in 2008. These programmes were divided into the following 8 genres according to an
internal BBC classification:

\begin{itemize}[noitemsep]
  \item Advice: Consumer, DIY and property shows.
  \item Children's: Including cartoons and educational shows.
  \item Comedy: Sit--coms and light entertainment shows.
  \item Competition: Quiz shows and other contest shows.
  \item Documentary: Including fly--on--the--wall shows.
  \item Drama: Soap operas and other serialised dramas.
  \item Events: Live events, sports and concerts.
  \item News: Broadcast news and current affair shows.
\end{itemize}

These genres are very heterogeneous, as the BBC classifies a large number of subgenres.
For instance, the ``Events'' genre covers music shows as well as live sports; or the
``Documentary'' genre covers nature documentaries as well as fly--on--the--wall shows.
Since the dataset contains all the BBC broadcasts from a single week, covering all genres, it is
a very complete scenario for the evaluation of background characterisation and genre identification techniques.

For the experiments, 285 shows were used for training and 47 shows were used for testing.
The number of shows and amount of time covered by each genre is presented in Table \ref{tab:time}. The selection of the test set was done with
the idea of providing equal coverage of genres and subgenres, with each genre represented by around 3 hours of broadcast time, except for
documentaries that have a larger representation due to the multiple subgenres existing. Shows in the test set were also classified depending whether
a previous instalment of the same show appeared in the training set, as this indicated whether
speakers and environments appearing in the test set also appeared in the training set. 28 of the 47 shows had previous instalments in the
training set, with the remaining 19 shows being unique instances of a show in the whole set.

\begin{table} [t,h]
\caption{\label{tab:time} {\it Distribution of shows by genre.}}
\centerline{
\begin{tabular}{|c|c|c|c|c|}
\hline
 &  \multicolumn{2}{|c|}{Train} & \multicolumn{2}{|c|}{Test} \\
\hline 
Genre & Shows & Time & Shows & Time \\
\hline
Advice & 34 & 24.5h. & 4 & 3.0h. \\
Children's & 45 & 18.5h. & 8 & 3.0h. \\
Comedy & 20 & 9.7h. & 6 & 3.2h. \\
Competition & 37 & 25.9h. & 6 & 3.3h. \\
Documentary & 41 & 29.8h. & 9 & 6.8h. \\
Drama & 19 & 14.4h. & 4 & 2.7h. \\
Events & 23 & 29.8h. & 5 & 4.3h. \\
News & 66 & 50.3h. & 5 & 2.0h. \\
\hline
Total & 285 & 203.0h.  & 47 & 28.3h.  \\
\hline
\end{tabular}}
\end{table}

While full transcriptions were not available for these shows, the close captioning subtitles that were broadcast with the shows
were available in order to train HMMs for ASR in an Hidden Markov Model Toolkit (HTK) setup \cite{HTK} using a lightly supervised
training process \cite{Lanchantin13}.
7 CMLLR asynchronous transformations were originally trained on a modified version of the
WSJCAM0 corpus \cite{WSJCAM0}, used in adaptation experiments \cite{Saz14}, containing 7 types of acoustic backgrounds: clean speech;
classical music; contemporary music; applause; cocktail party noise; traffic noise and
wildlife noise. These transformations were asynchronously retrained in the BBC dataset to represent the different acoustic disturbances
present in this data. The asynchronous alignment required to extract background--tracking features was also performed based on the
existing subtitles with $T=7$ and $P=100$, so a 7-dimensional feature vector was extracted from each second of the input audio.

An illustration of the output of the background--tracking feature extraction can be seen in the images in Figure \ref{fig:samples}. These images
visualise the 7--dimensional feature vectors extracted, as explained earlier in the Section. These samples represent 4 periods of one minute
(60 frames) from 4 different shows. The values of each of the 7 dimensions
are represented by the size of the 7 coloured bars in each frame.
Figure \ref{fig:news} is one minute in a broadcast news programme, where the background changes
from music to street noise to clean studio and ends with street noise. Figure \ref{fig:events} is one minute in a music event show, where the music changes from
rock music to solo singing and then to instrumental rock music. Figure \ref{fig:documentary} is one minute in a historical documentary show,
that starts with bell sounds, followed by a period of music, another period of clean speech and finishes with sounds of seaside and birds.
Finally, Figure \ref{fig:comedy} is one minute in a light entertainment show that mixes speech with long bursts of laughter.

\section{Results}
\label{sec:results}

\begin{figure}[t]
\centering
\epsfig{figure=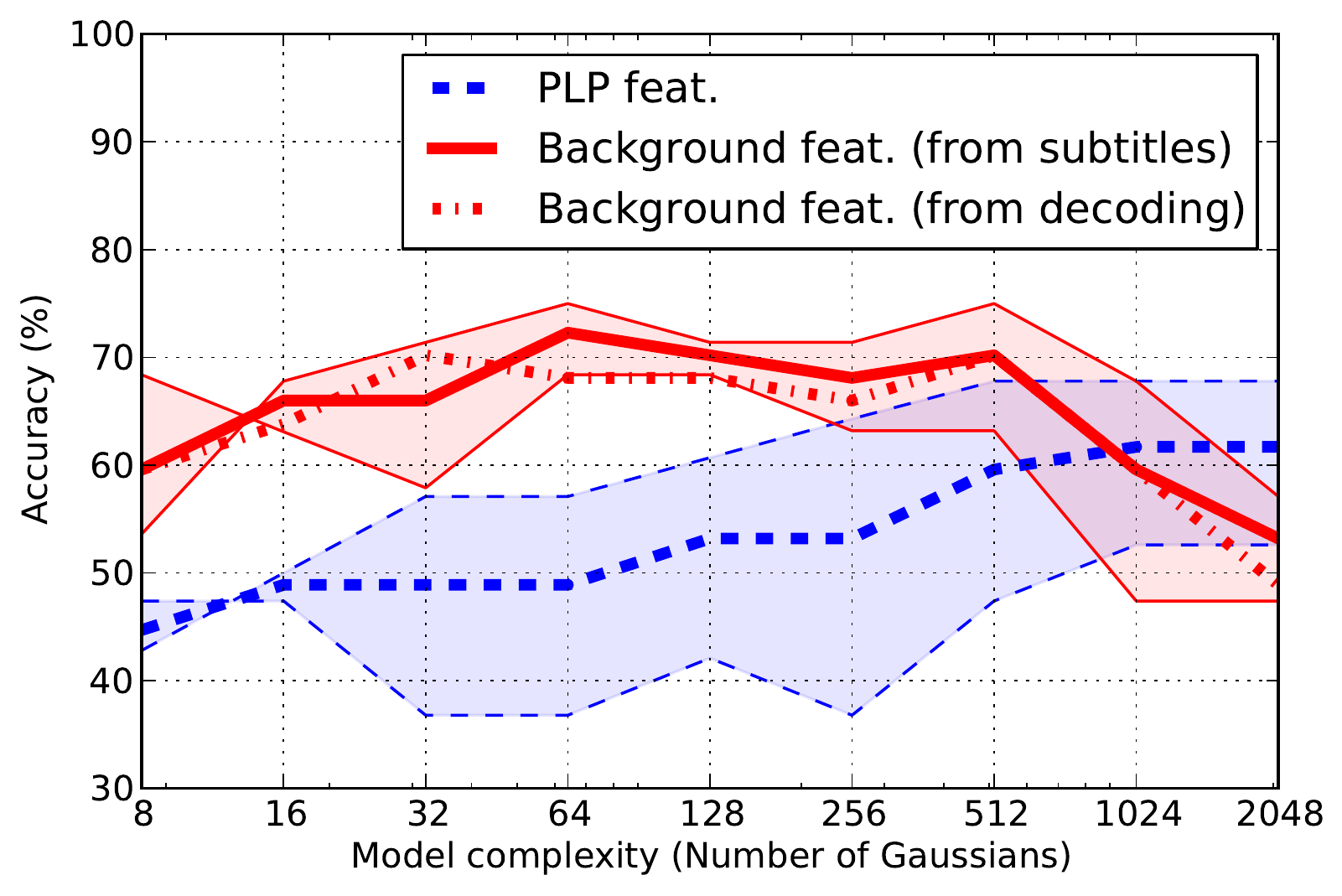,width=80mm}
\caption{{Accuracy in genre identification PLP and background--tracking features with GMM classifiers (Thicker lines represent global accuracy, thinner lines
represent accuracy for repeated and non-repeated shows).}}
\label{fig:plp-user}
\end{figure}

The first set of experiments were designed to evaluate the performance of the proposed background--tracking features
compared to short--term features in the genre identification task. Genre--based GMMs were trained with the feature vectors
extracted from all the shows in the training set belonging to each genre. A set of GMMs was trained with 
13--dimenstional PLP features extracted every 10 ms. and another set with 7--dimensional background--tracking features
extracted every second. First and second derivatives were also computed and added to the feature vectors, for a total of 39 dimensions
in the PLP features and 21 dimensions in the background--tracking features. The background--tracking features were tested on two conditions,
the first one assuming that the subtitles of the shows in the test set were available for the alignment, and the second one using the transcription
provided by the ASR system to do the alignment. The classification of the genre for each show in the test set was
done by selecting the GMM that maximised the overall likelihood of all the input frames in the test show.
The results in terms of accuracy (number of correctly classified shows divided by the total number of test shows)
for different number of Gaussians in the GMMs for both types of features are presented in Figure \ref{fig:plp-user}.

Background--tracking features outperformed PLPs in this task. While the proposed features
achieved up to 72.4\% accuracy, PLPs only reached 61.7\% accuracy. In further analysis, PLPs required a higher number of Gaussians (up to 1,024 and 2,048) to achieve their
best performance, while background--tracking features required less model complexity.
This was due to the long--term nature of the background--tracking features, which were extracted
every second, instead of every 10 milliseconds.
While a total of 73,528,233 frames were available for training the PLP GMMs, only 730,621 were available with the background--tracking features.
Figure \ref{fig:plp-user} also shows that there was little difference between using the subtitles or the decoding transcripts to extract the background--tracking features
in the test shows, indicating that the feature extraction process was robust to the use of noisy transcriptions in the asynchronous alignment.
Following this, all further experiments
were based on the alignment to the subtitles.

The final element for analysis is presented in thinner lines around the main lines in Figure \ref{fig:plp-user}. These lines
mark the accuracies achieved in shows that have previous instalments in the training set
and the accuracy achieved in the rest of the shows. PLP features
presented a larger spread (represented by the shaded area in the Figure) between these two types of shows, 15\% to 20\% difference in absolute accuracy across most of the range of GMM sizes,
while background--tracking exhibited lower difference, 5\% to 10\% maximum. For shows with previous
episodes in the training set, PLP features achieved 67.8\% accuracy, narrowing the gap to the 75.0\% obtained with background--tracking features
for the same shows. However, for the rest of the shows, PLPs only reached 52.6\% accuracy, while background--tracking features
reached a more robust 68.4\%.
This pointed out
how short--term features were more sensitive to the presence of known speakers and environments in the training set.

Afterwards, more advanced classifiers were evaluated using background--tracking features.
Two experiments were set to study two aspects of classification: Modelling of temporal changes and discriminative methods.
The first classifier used were HMMs, which are generative classifiers like GMMs,
but, unlike GMMs, they also model temporal transitions among hidden states existing in the input data.
For these experiments, HMMs with 8 states were found to provide the best performance and were, subsequently, used. The Gaussian components in each state and the
transition probabilities among states were learnt using a Maximum Likelihood (ML) \cite{Dempster77}
approach from all the input feature vectors from the shows in the training data. The selection of the
genre for each test show was also done maximising the likelihood.

The second classifier used at this stage were Support Vector Machines (SVM) \cite{Cortes95}. SVMs are widely used discriminative classifiers
and had been previously used in the genre identification task \cite{Ekenel13}. In these experiments, the inputs to the SVM classifier were supervectors
obtained by concatenating the Gaussian means of show--based GMMs trained via Maximum A Posteriori (MAP) adaptation \cite{Gauvain94}.
Gaussian--kernel SVMs were trained \cite{Joachims99} for each genre to classify
whether shows belonged or not to that genre. The final decision for each test show was made for the genre whose SVM gave the best score from
all the genre--based SVMs.

\begin{figure}[t]
\centering
\epsfig{figure=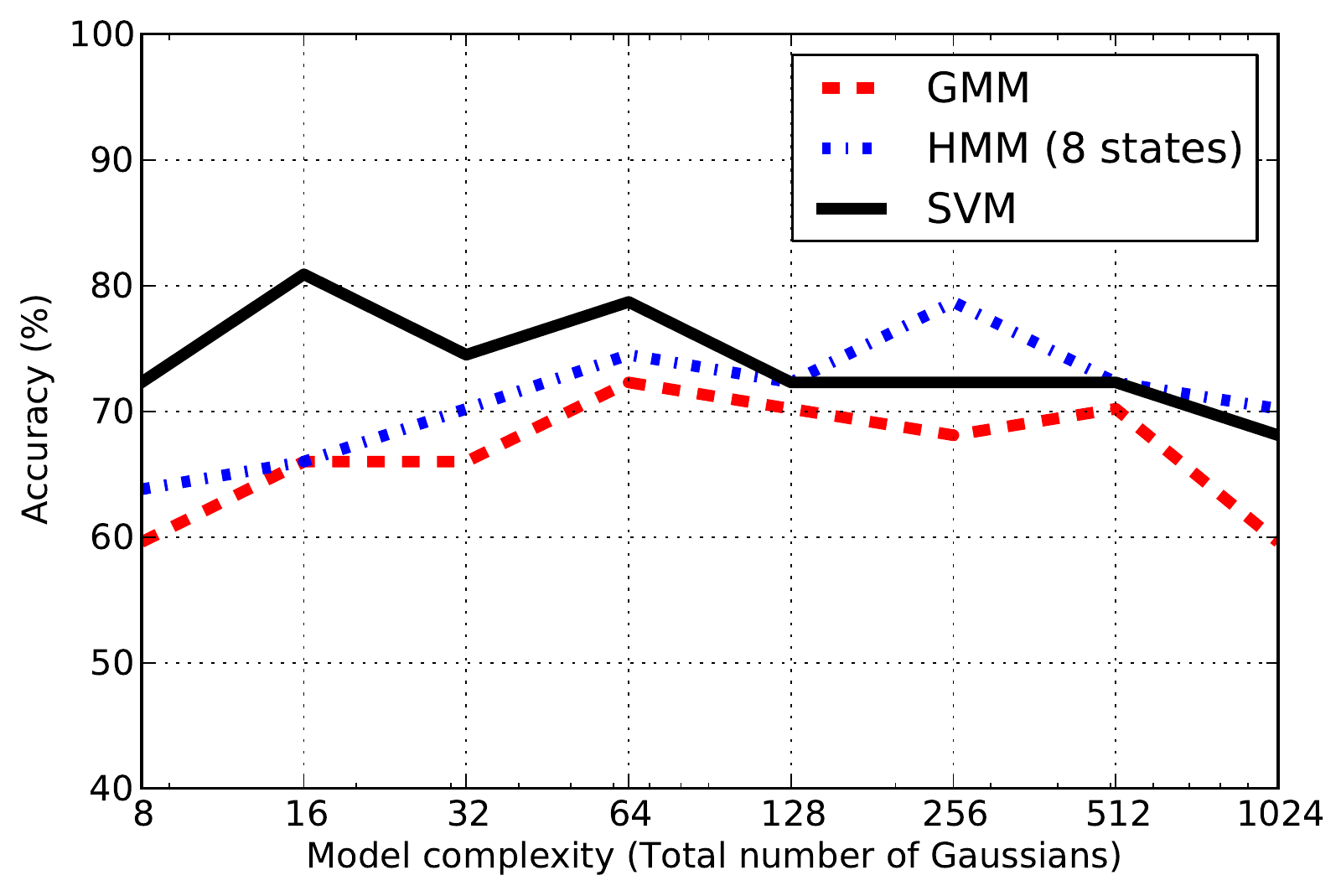,width=80mm}
\caption{{Accuracy in genre identification of GMM, HMM and SVM classifiers using background--tracking features.}}
\label{fig:gmm-hmm}
\vspace{-4mm}
\end{figure}

The results of the GMM, HMM and SVM classifiers are shown in Figure \ref{fig:gmm-hmm} for different values of model complexity.
They showed
that both HMMs and SVMs outperformed GMMs. The best result for HMMs, 78.7\% accuracy, was achieved with a total model complexity of
256 Gaussians (8 states with 32 Gaussians each); while the best result for SVMs, 80.9\% accuracy, was achieved with a lower model complexity, only 16 Gaussians.

\begin{figure}[t]
\centering
\epsfig{figure=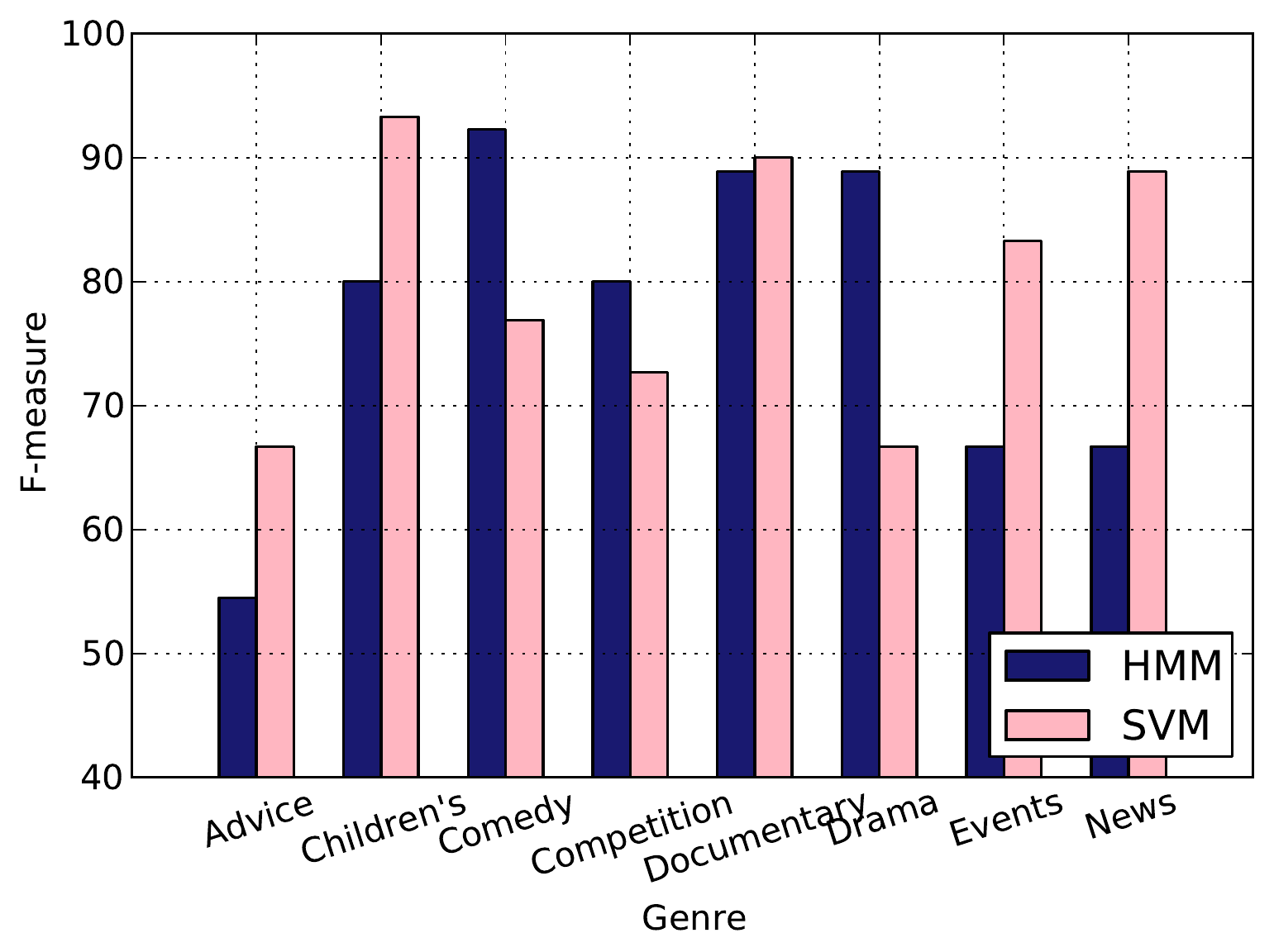,width=79mm}
\caption{{F--measure for HMM and SVM classifiers.}}
\label{fig:f}
\vspace{-4mm}
\end{figure}

To evaluate the identification abilities of the proposed systems, the F--measure of the two best HMM and SVM systems
are presented in Figure \ref{fig:f} for each genre. The F-measure, defined as the harmonic mean of precision and recall for each class,
allows to evaluate the accuracy and specificity of a classifier.
Figure \ref{fig:f} shows that SVMs performed better identifying the ``Advice'', ``Children's'', ``Events'' and
``News'' genres, while HMMs outperformed SVMs in the ``Comedy'', ``Competition'' and ``Drama'' genres.

Finally,
system combination based on the confidence scores given by the best HMM and SVM systems was performed \cite{Silva10}.
System combination has traditionally been proposed as a solid way of exploiting the outputs of different classifiers with different properties; in this
task, the modelling of dynamics given by HMMs and the discriminative modelling provided by SVMs.
The confidence of the HMM classifier was based on the likelihood score of the decided HMM; while the confidence score
of the SVM classifier was based on the distance score provided by
the decided SVM, both normalised to the range of $[0,1]$.
When both systems provided the same hypothesis, this was accepted straightaway; but when they disagreed, the output of the system with highest confidence
was selected. The result of the combination of both systems in terms of global accuracy was 83.0\%.



\vspace{-1mm}
\section{Conclusions}
\label{sec:conclusion}
\vspace{-1mm}

The proposed background--tracking features have shown, through a range of different classifiers, that they can provide robust results in the
task of genre identification of broadcast shows. While, in absolute terms, the use of acoustic and video features
has been reported to provide better performance \cite{Montagnuolo07,Montagnuolo09,Ekenel13}, the results are
very promising when compared with previous results using only acoustic features. Furthermore, some types of
broadcasts, such as radio or podcasts, do not have video and rely only on the acoustics for classification.
Future work will have to see these novel
acoustic features merged with state--of--the--art video features to compare with the best performing systems
in this task.

The experiments have also shown that the use of long--term features outperforms usual short--term features
in tasks that require an acoustic characterisation of the background.
Features like PLPs or MFCCs have great classification capabilities in speech but fail to generalise
well, as shown by \cite{Ekenel13} in their comparison of different datasets, because they mostly describe the
phonemes or speakers in the audio. Long--term background--based features provide a more comprehensive
description of the acoustic conditions of broadcasts, and are less sensitive to the recurring presence or not of the
same speakers and environments.

There are many other tasks where the background--tracking features could be exploited. In the future,
these features can be used to automatically split complete shows or videos into homogeneous
segments with a similar acoustic background. These segments could be clustered by similarity and then
used to let users browse and link segments with a similar acoustic background. From the point of view of
speech technologies, it is needed to explore how these features can be used in ASR tasks in noisy conditions.
Background--tracking features could be used to adapt or compensate to background noises and disturbances,
even in the case when the background changes asynchronously, enhancing ASR performance.

\bibliographystyle{IEEEbib}

\end{document}